\documentclass[aps,twocolumn,showpacs]{revtex4}
\usepackage{epsfig}
\usepackage{longtable}
\newcommand{\nc}{\newcommand}
\nc{\rnc}{\renewcommand}

\nc{\beq}{\begin{equation}}
\nc{\eeq}{\end{equation}}
\nc{\bea}{\begin{eqnarray}}
\nc{\eea}{\end{eqnarray}}
\nc{\ba}{\begin{array}}
\nc{\ea}{\end{array}}
\nc{\bpi}{\begin{picture}}
\nc{\epi}{\end{picture}}
\nc{\nn}{\nonumber}

\nc{\p}{\partial}
\nc{\f}[2]{\frac{#1}{#2}}
\nc{\od}{{\cal O}}
\nc{\ra}{\rightarrow}
\nc{\Rcal}{{\cal R}}
\nc{\uh}{\hat{u}}

\nc{\al}{\alpha}
\nc{\be}{\beta}
\nc{\de}{\delta}
\nc{\om}{\omega}
\nc{\ze}{\zeta}
\nc{\De}{\Delta}
\nc{\Si}{\Sigma}

\begin{document}

\bibliographystyle{apsrev}

\title{Bose-Einstein Condensation Temperature of Homogenous Weakly
Interacting\\
Bose Gas in Variational Perturbation Theory Through Seven Loops}

\author{Boris Kastening}
\affiliation{Institut f\"ur Theoretische Physik\\
Freie Universit\"at Berlin\\ Arnimallee 14\\ D-14195 Berlin\\ Germany\\
{\tt boris.kastening@physik.fu-berlin.de}}

\date{September 2003}

\begin{abstract}
The shift of the Bose-Einstein condensation temperature for a homogenous
weakly interacting Bose gas in leading order in the scattering
length $a$ is computed for given particle density $n$.
Variational perturbation theory is used to resum the corresponding
perturbative series for $\De\langle\phi^2\rangle/Nu$ in a classical
three-dimensional scalar field theory with coupling $u$ and where the
physical case of $N=2$ field components is generalized to arbitrary $N$.
Our results for $N=1,2,4$ are in agreement with recent Monte-Carlo
simulations; for $N=2$, we obtain $\De T_c/T_c=1.27\pm0.11\,an^{1/3}$.
We use seven-loop perturbative coefficients, extending earlier work by one
loop order.
\end{abstract}

\maketitle

\section{Introduction}
The recent advent of experimental realizations of Bose-Einstein
condensation (BEC) in dilute gases has been followed by an intense
revival of theoretical investigation of the subject.
The diluteness of the gas allows for an approximation where the
interactions of the bosons are described by just one
parameter, the $s$-wave scattering length $a$ corresponding to the
two-particle interaction potential.
For many questions of interest, mean-field theory may be applied as
a further approximation.
An exception is the physics around the condensation temperature $T_c$.
Perturbation theory (PT) breaks down close to the transition due to
infrared (IR) divergences.
The phase transition is second-order and the interactions change the
universality class from Gaussian to that of a three-dimensional
O(2)-symmetric scalar field theory.

In this work, we are interested in the shift of the condensation
temperature of a homogenous gas away from its ideal-gas value 
\beq
\label{t0n0}
T_0=\f{2\pi}{m}\left[\f{n}{\ze(3/2)}\right]^{2/3},
\eeq
where $m$ is the mass of the bosons, $n$ their number density and we
work throughout in units where $k_B=\hbar=1$.
There have been many attempts to solve this problem
\cite{history,Sto,other,GrCeLa,HoGrLa,HoKr,BaBlHoLaVa1,BaBlZi,
WiIlKr,ArTo,Al,SCPiRa1,KaPrSv,ArMo,HoBaBlLa,ArMoTo,BaBlHoLaVa2,SCPiRa2,
KnPiRa,BrRa1,BrRa2,Kl1,Ka}, but the non-perturbative nature of the physics
around the phase transition makes the problem highly non-trivial.
More precisely, we ask what the value of the constant $c_1$ is
in the diluteness expansion \cite{BaBlHoLaVa1,BaBlZi,HoBaBlLa}
\beq
\label{deltatexp}
\f{\De T_c}{T_0}=c_1an^{1/3}+[c_2'\ln(an^{1/3})+c_2''](an^{1/3})^2+\cdots
\eeq
While $c_2'$ can be evaluated exactly, $c_2''$ involves a non-perturbative
calculation similar to the one needed for $c_1$.
For results for both $c_2'$ and $c_2''$, the reader is referred to
Ref.\ \cite{ArMoTo}.

Results for the constant $c_1$ range from $-0.93$ \cite{WiIlKr}
to $4.7$ \cite{Sto} as has been summarized for instance in
Refs.\ \cite{ArMo,BaBlHoLaVa2} and the review \cite{An}, see also
Fig.\ \ref{fig3} below.
In \cite{BaBlHoLaVa1} it was realized that $c_1$ is generated
exclusively by long-wavelength fluctuations and that therefore
three-dimensional field theory is a convenient tool for the
determination of $c_1$.
Subsequently, schemes have been set up that are geared towards
exclusively determining $c_1$ without having to disentangle
the different contributions in (\ref{deltatexp}).
It appears that the most reliable results so far are obtained by
Monte-Carlo (MC) simulations \cite{KaPrSv,ArMo}, as we will explain
in Sec.\ \ref{discussion}.

The strategy employed here is to expand $c_1$ in a perturbative
series, which is subsequently resummed to obtain the physical
non-perturbative limit.
While naive schemes such as Pad\'e approximants  may slowly converge
towards the correct value, a scheme suited particularly well for critical
phenomena is Kleinert's field theoretic variational perturbation theory
(VPT, see \cite{Kl2,Kl3,Kl4} and Chapters 5 and 19 of the textbooks
\cite{pibook} and \cite{phi4book}, respectively).
Improving PT by a variational principle goes back at
least to \cite{Yu}.
Several other determinations of $c_1$ use variationally resummed
PT employing the linear $\de$ expansion (LDE)
\cite{SCPiRa1,SCPiRa2,KnPiRa,BrRa1,BrRa2}.
These were criticized in Refs.\ \cite{Kl1,HaKl}.
In \cite{Kl1} VPT was used through five loops.
In \cite{Ka}, we extended the calculation of \cite{Kl1} to six loops,
but with a different treatment of the one-loop term, which is absent
in \cite{Ka}:
The first non-zero coefficient arises only at three loops.
In consequence, we obtained a result about 30\% larger than that of
Ref.\ \cite{Kl1}.
We comment on the issue at the end of Sec.\ \ref{fieldtheory} below.
Here we extend the work in \cite{Ka} to seven loops, which reduces
the error bar of the final result.
We also provide many details and a comparison with MC simulation results
not only for the case of $N=2$ real field components, relevant for BEC,
but also for $N=1$ and $N=4$, which were treated with MC methods in
\cite{Su}.

This work is organized as follows.
In Sec.\ \ref{fieldtheory} we set up the field-theoretic framework
and repeat arguments why, for the determination of $c_1$, we may
work entirely in three dimensions, ignoring the (imaginary) time
direction. 
In Sec.\ \ref{pt} we discuss which perturbative series we need to resum
to obtain $c_1$ and provide perturbative coefficients through seven loops.
In Sec.\ \ref{resummation} the resummation procedure is described and
carried out and the results are presented.
Sec.\ \ref{discussion} concludes with a discussion.

\section{Field-theoretic Considerations}
\label{fieldtheory}
The theoretical basis for the description of equilibrium quantities of
a gas of identical spin-$0$ bosons in the grand canonical ensemble is
a non-relativistic $(3+1)$-dimensional field theory in imaginary time
$\tau$, given by the Euclidean action
\beq
S_{3+1}
\!=\!\!
\int_0^\be\!\!\!d\tau\!\!\int\!\! d^3x\bigg[\psi^*\!\left(\f{\p}{\p\tau}
{-}\f{1}{2m}\nabla^2{-}\mu\right)\psi
+\f{2\pi a}{m}(\psi^*\psi)^2\bigg],
\eeq
where $\mu$ is the chemical potential.
This assumes that the momenta of the particles are small enough
so that their interaction potential is well described by just one
parameter, the $s$-wave scattering length $a$.
It also assumes that three- and more particle interactions are rare,
i.e.\ that the gas is dilute.
As we shall see below, the leading perturbative contribution to
$\De T_c$ is $\propto a^2$.
A consequence is that the leading contribution to $\De T_c$ arises
exclusively from terms that are infrared divergent in the framework
of PT.
The leading IR divergent diagrams involve, however, only propagators
with zero Matsubara frequencies.
Put another way, a diagram with a propagator with a non-zero Matsubara
frequency has the same degree of IR divergence as a diagram with no
non-zero Matsubara frequencies but less powers of $a$ and is therefore
not contributing in leading order to $\De T_c$.
This is why, for the determination of $c_1$, we can set all Matsubara
frequencies to zero or, equivalently, work with a three-dimensional
field theory \cite{BaBlHoLaVa1}.
Denoting the zero-Matsubara modes by $\psi_0$, we may define the fields
and parameters of this theory by
$\psi_0=\sqrt{mT}(\phi_1+i\phi_2)$, $r_{\rm bare}=-2m\mu$, $u=48\pi amT$
and obtain the three-dimensional Euclidean action
\beq
\label{s3}
S_3=\int d^3x\left[\f{1}{2}|\nabla\phi|^2
+\f{r_{\rm bare}}{2}\phi^2+\f{u}{24}(\phi^2)^2\right].
\eeq
For the determination of quantities that are not governed by the
leading IR divergences, one may still use a three-dimensional field
theory which, however, is obtained by a more complicated matching
procedure, which introduces corrections to the above relations of
the parameters of the $(3+1)$-dimensional and the three-dimensional
theory and also necessitates the inclusion of more interaction terms.
This matching is described in detail in \cite{ArMoTo}.

It is convenient to generalize the model (\ref{s3}) to an O($N$)
field theory, where $\phi=(\phi_1,\ldots,\phi_N)$,
$\phi^2\equiv\phi_a\phi_a$.
This allows to make contact with the exactly known large-$N$ result
for $c_1$ \cite{BaBlZi} and with the MC results for $N=1$ and $N=4$
\cite{Su}.

From the form of the ideal-gas result (\ref{t0n0}) one obtains, in
leading order, the relation \cite{BaBlHoLaVa1}
\beq
\label{detc}
\f{\De T_c}{T_0}=-\f{2}{3}\f{\De n}{n},
\eeq
where $\De T_c$ is the shift of the condensation temperature for
fixed $n$ and $\De n$ is the shift of the critical particle density
for fixed condensation temperature.
In our field-theoretic setup, $\De n$ is given by
\beq
\label{deltanc}
\De n=\De\langle\psi^*\psi\rangle=mT\De\langle\phi^2\rangle.
\eeq
Combining (\ref{t0n0}), (\ref{deltatexp}), (\ref{detc}) and (\ref{deltanc}),
$c_1$ is given by
\beq
\label{c1def}
c_1=\al\left.\f{\De\langle\phi^2\rangle}{Nu}\right|_{\rm crit.}
\eeq
with $\al=-256\pi^3/[\ze(3/2)]^{4/3}\approx-2206.19$ and the remaining
task is to compute the critical limit of
\beq
\label{deltaphi2}
\f{\De\langle\phi^2\rangle}{Nu}
=\f{1}{u}\int\f{d^3p}{(2\pi)^3}[G(p)-G_0(p)],
\eeq
where $G$ and $G_0$ are the interacting and the free propagator,
respectively.
The theory (\ref{s3}) is superrenormalizable, which means that only
a finite number of primitive diagrams are divergent.
Since the divergence in the free energy is of no interest for us
here, the only divergences we have to cancel appear in the self-energy
$\Si(p)$.
They are, however, $p$-independent and can be absorbed into $r_{\rm bare}$
by defining the renormalized quantity $r=r_{\rm bare}-\Si(0)$.
This renormalization scheme has two major advantages:
(i) It allows to work without any regulator to obtain renormalized results.
(ii) The critical limit is easily identified as $r\ra0$, since
the full  propagator reads now $G(p)=1/\{p^2+r-[\Si(p)-\Si(0)]\}$.
For the free propagator follows $G_0(p)=1/(p^2+r)$ \cite{BrRa1}.

Another popular scheme, used in \cite{SCPiRa1,SCPiRa2,KnPiRa,Kl1,HaKl},
is to take $G_0(p)=1/p^2$.
This generates an unnatural non-zero one-loop contribution to
$\De\langle\phi^2\rangle$, even in the absence of interactions.
Although this contribution vanishes as $r\ra0$, it strongly
influences resummation and is responsible for the small value of
$c_1=0.92\pm0.13$ \cite{Kl1res} and partial absence of real variational
solutions in Ref.\ \cite{Kl1}.

\section{Perturbative Series for \boldmath$c_1$}
\label{pt}
Dimensional analysis shows that PT organizes
$\De\langle\phi^2\rangle/Nu$ in a power series of $u/r^{1/2}$.
It is convenient to define $c_1$ as in (\ref{c1def}) but as a function
of $u/r^{1/2}$, without yet taking the critical limit.
Specifically, we write
\beq
\label{pertsum}
c_1(u_r)=\al\sum_{l=1}^\infty a_lu_r^{l-2},~~~~~~
u_r\equiv\f{Nu}{4\pi r^{1/2}}.
\eeq
As we will see in Sec.\ \ref{resummation}, we can compute $c_1(u_r)$
through resummation all the way from the perturbative region around
$u_r=0$ to $u_r\ra\infty$, where it assumes the value which we need
for the determination of $\De T_c$.
The specific choice of the variable $u_r$ allows to go smoothly
to the $N\ra\infty$ limit with finite perturbative coefficients $a_l$.
Comparison with the large-$N$ limit is useful since $c_1$ has been
determined exactly for this case in \cite{BaBlZi}.
The successful application of VPT to the large-$N$ limit has been
demonstrated in \cite{Ka}.

The perturbative series can be represented by Feynman diagrams and
then becomes a loop expansion.
We have constructed the relevant diagrams through seven loops using
the recursive methods of \cite{recrel} and collect them in
Table \ref{diagrams}.
\begin{table*}
\begin{tabular}{c}
\includegraphics[width=1.5cm,angle=0]{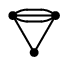}\\
3-1
\end{tabular}
\begin{tabular}{c}
\includegraphics[width=1.5cm,angle=0]{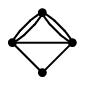}\\
4-1
\end{tabular}
\begin{tabular}{c}
\includegraphics[width=1.5cm,angle=0]{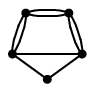}\\
5-1
\end{tabular}
\begin{tabular}{c}
\includegraphics[width=1.5cm,angle=0]{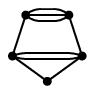}\\
5-2
\end{tabular}
\begin{tabular}{c}
\includegraphics[width=1.5cm,angle=0]{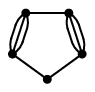}\\
5-3
\end{tabular}
\begin{tabular}{c}
\includegraphics[width=1.5cm,angle=0]{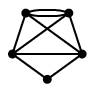}\\
5-4
\end{tabular}
\begin{tabular}{c}
\includegraphics[width=1.5cm,angle=0]{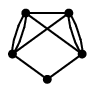}\\
5-5
\end{tabular}
\begin{tabular}{c}
\includegraphics[width=1.5cm,angle=0]{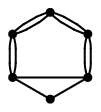}\\
6-1
\end{tabular}
\begin{tabular}{c}
\includegraphics[width=1.5cm,angle=0]{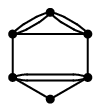}\\
6-2
\end{tabular}
\begin{tabular}{c}
\includegraphics[width=1.5cm,angle=0]{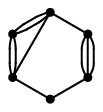}\\
6-3
\end{tabular}
\begin{tabular}{c}
\includegraphics[width=1.5cm,angle=0]{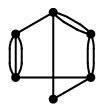}\\
6-4
\end{tabular}
\begin{tabular}{c}
\includegraphics[width=1.5cm,angle=0]{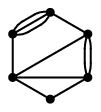}\\
6-5
\end{tabular}
\begin{tabular}{c}
\includegraphics[width=1.5cm,angle=0]{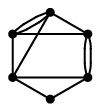}\\
6-6
\end{tabular}
\begin{tabular}{c}
\includegraphics[width=1.5cm,angle=0]{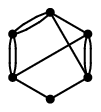}\\
6-7
\end{tabular}
\begin{tabular}{c}
\includegraphics[width=1.5cm,angle=0]{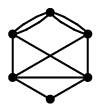}\\
6-8
\end{tabular}
\begin{tabular}{c}
\includegraphics[width=1.5cm,angle=0]{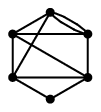}\\
6-9
\end{tabular}
\begin{tabular}{c}
\includegraphics[width=1.5cm,angle=0]{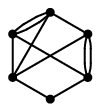}\\
6-10
\end{tabular}
\begin{tabular}{c}
\includegraphics[width=1.5cm,angle=0]{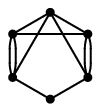}\\
6-11
\end{tabular}
\begin{tabular}{c}
\includegraphics[width=1.5cm,angle=0]{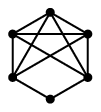}\\
6-12
\end{tabular}
\begin{tabular}{c}
\includegraphics[width=1.5cm,angle=0]{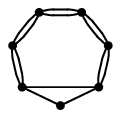}\\
7-1
\end{tabular}
\begin{tabular}{c}
\includegraphics[width=1.5cm,angle=0]{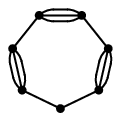}\\
7-2
\end{tabular}
\begin{tabular}{c}
\includegraphics[width=1.5cm,angle=0]{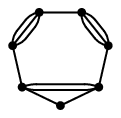}\\
7-3
\end{tabular}
\begin{tabular}{c}
\includegraphics[width=1.5cm,angle=0]{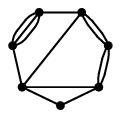}\\
7-4
\end{tabular}
\begin{tabular}{c}
\includegraphics[width=1.5cm,angle=0]{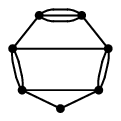}\\
7-5
\end{tabular}
\begin{tabular}{c}
\includegraphics[width=1.5cm,angle=0]{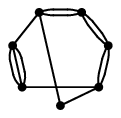}\\
7-6
\end{tabular}
\begin{tabular}{c}
\includegraphics[width=1.5cm,angle=0]{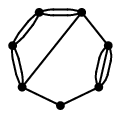}\\
7-7
\end{tabular}
\begin{tabular}{c}
\includegraphics[width=1.5cm,angle=0]{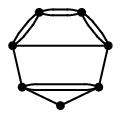}\\
7-8
\end{tabular}
\begin{tabular}{c}
\includegraphics[width=1.5cm,angle=0]{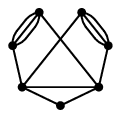}\\
7-9
\end{tabular}
\begin{tabular}{c}
\includegraphics[width=1.5cm,angle=0]{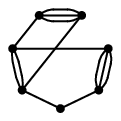}\\
7-10
\end{tabular}
\begin{tabular}{c}
\includegraphics[width=1.5cm,angle=0]{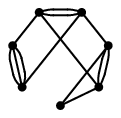}\\
7-11
\end{tabular}
\begin{tabular}{c}
\includegraphics[width=1.5cm,angle=0]{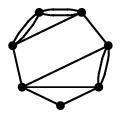}\\
7-12
\end{tabular}
\begin{tabular}{c}
\includegraphics[width=1.5cm,angle=0]{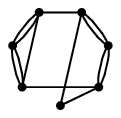}\\
7-13
\end{tabular}
\begin{tabular}{c}
\includegraphics[width=1.5cm,angle=0]{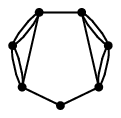}\\
7-14
\end{tabular}
\begin{tabular}{c}
\includegraphics[width=1.5cm,angle=0]{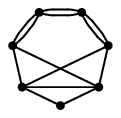}\\
7-15
\end{tabular}
\begin{tabular}{c}
\includegraphics[width=1.5cm,angle=0]{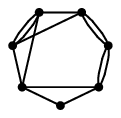}\\
7-16
\end{tabular}
\begin{tabular}{c}
\includegraphics[width=1.5cm,angle=0]{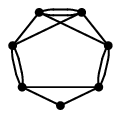}\\
7-17
\end{tabular}
\begin{tabular}{c}
\includegraphics[width=1.5cm,angle=0]{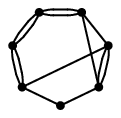}\\
7-18
\end{tabular}
\begin{tabular}{c}
\includegraphics[width=1.5cm,angle=0]{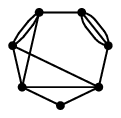}\\
7-19
\end{tabular}
\begin{tabular}{c}
\includegraphics[width=1.5cm,angle=0]{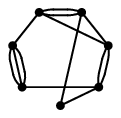}\\
7-20
\end{tabular}
\begin{tabular}{c}
\includegraphics[width=1.5cm,angle=0]{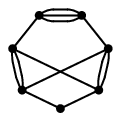}\\
7-21
\end{tabular}
\begin{tabular}{c}
\includegraphics[width=1.5cm,angle=0]{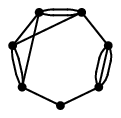}\\
7-22
\end{tabular}
\begin{tabular}{c}
\includegraphics[width=1.5cm,angle=0]{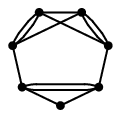}\\
7-23
\end{tabular}
\begin{tabular}{c}
\includegraphics[width=1.5cm,angle=0]{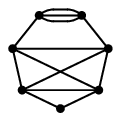}\\
7-24
\end{tabular}
\begin{tabular}{c}
\includegraphics[width=1.5cm,angle=0]{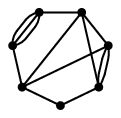}\\
7-25
\end{tabular}
\begin{tabular}{c}
\includegraphics[width=1.5cm,angle=0]{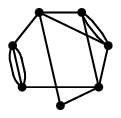}\\
7-26
\end{tabular}
\begin{tabular}{c}
\includegraphics[width=1.5cm,angle=0]{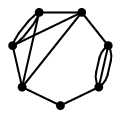}\\
7-27
\end{tabular}
\begin{tabular}{c}
\includegraphics[width=1.5cm,angle=0]{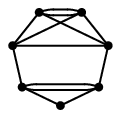}\\
7-28
\end{tabular}
\begin{tabular}{c}
\includegraphics[width=1.5cm,angle=0]{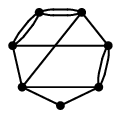}\\
7-29
\end{tabular}
\begin{tabular}{c}
\includegraphics[width=1.5cm,angle=0]{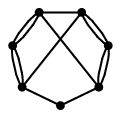}\\
7-30
\end{tabular}
\begin{tabular}{c}
\includegraphics[width=1.5cm,angle=0]{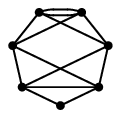}\\
7-31
\end{tabular}
\begin{tabular}{c}
\includegraphics[width=1.5cm,angle=0]{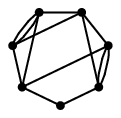}\\
7-32
\end{tabular}
\begin{tabular}{c}
\includegraphics[width=1.5cm,angle=0]{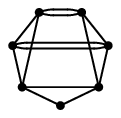}\\
7-33
\end{tabular}
\begin{tabular}{c}
\includegraphics[width=1.5cm,angle=0]{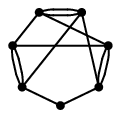}\\
7-34
\end{tabular}
\begin{tabular}{c}
\includegraphics[width=1.5cm,angle=0]{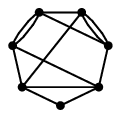}\\
7-35
\end{tabular}
\begin{tabular}{c}
\includegraphics[width=1.5cm,angle=0]{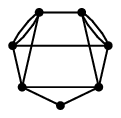}\\
7-36
\end{tabular}
\begin{tabular}{c}
\includegraphics[width=1.5cm,angle=0]{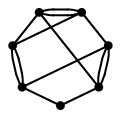}\\
7-37
\end{tabular}
\begin{tabular}{c}
\includegraphics[width=1.5cm,angle=0]{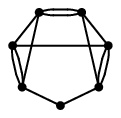}\\
7-38
\end{tabular}
\begin{tabular}{c}
\includegraphics[width=1.5cm,angle=0]{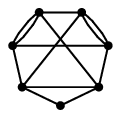}\\
7-39
\end{tabular}
\begin{tabular}{c}
\includegraphics[width=1.5cm,angle=0]{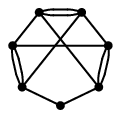}\\
7-40
\end{tabular}
\begin{tabular}{c}
\includegraphics[width=1.5cm,angle=0]{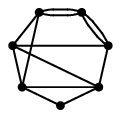}\\
7-41
\end{tabular}
\begin{tabular}{c}
\includegraphics[width=1.5cm,angle=0]{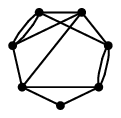}\\
7-42
\end{tabular}
\begin{tabular}{c}
\includegraphics[width=1.5cm,angle=0]{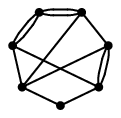}\\
7-43
\end{tabular}
\begin{tabular}{c}
\includegraphics[width=1.5cm,angle=0]{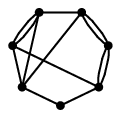}\\
7-44
\end{tabular}
\begin{tabular}{c}
\includegraphics[width=1.5cm,angle=0]{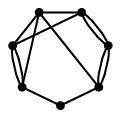}\\
7-45
\end{tabular}
\begin{tabular}{c}
\includegraphics[width=1.5cm,angle=0]{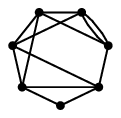}\\
7-46
\end{tabular}
\begin{tabular}{c}
\includegraphics[width=1.5cm,angle=0]{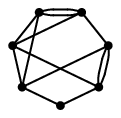}\\
7-47
\end{tabular}
\begin{tabular}{c}
\includegraphics[width=1.5cm,angle=0]{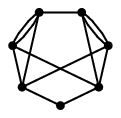}\\
7-48
\end{tabular}
\begin{tabular}{c}
\includegraphics[width=1.5cm,angle=0]{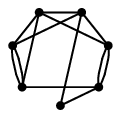}\\
7-49
\end{tabular}
\begin{tabular}{c}
\includegraphics[width=1.5cm,angle=0]{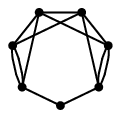}\\
7-50
\end{tabular}
\begin{tabular}{c}
\includegraphics[width=1.5cm,angle=0]{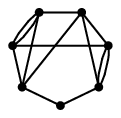}\\
7-51
\end{tabular}
\begin{tabular}{c}
\includegraphics[width=1.5cm,angle=0]{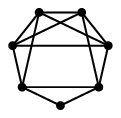}\\
7-52
\end{tabular}
\begin{tabular}{c}
\includegraphics[width=1.5cm,angle=0]{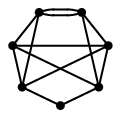}\\
7-53
\end{tabular}
\begin{tabular}{c}
\includegraphics[width=1.5cm,angle=0]{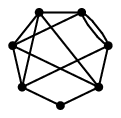}\\
7-54
\end{tabular}
\begin{tabular}{c}
\includegraphics[width=1.5cm,angle=0]{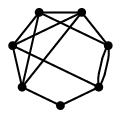}\\
7-55
\end{tabular}
\begin{tabular}{c}
\includegraphics[width=1.5cm,angle=0]{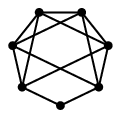}\\
7-56
\end{tabular}
\caption{\label{diagrams}
 Diagrams through seven loops.
The lines represent free propagators $\de_{ab}/(p^2+r)$. The vertices
connecting four lines represent the interaction
$-u(\de_{ab}\de_{cd}+\de_{ac}\de_{bd}+\de_{ad}\de_{bc})/3$.
The vertex connecting only two lines is an insertion $\de_{ab}$
connecting two free propagators.
Indices run from $1$ to $N$.}
\end{table*}

The subtraction of the zero-momentum part of the self-energy in the
full propagator has to be performed recursively for all subdiagrams.
Representing this recursive subtraction by an operator $\cal R$ \cite{Ka},
the perturbative coefficient at $L$ loops is given by
\beq
a_Lu_r^{L-2}=\f{1}{Nu}\sum_{k=1}^{n_L}{\cal R}D_{L-k},
\eeq
where $D_{L-k}$ is the $k$-th $L$-loop diagram and the first few $n_L$
are given in Table \ref{nd}.
Here we count only diagrams surviving the subtractions of $\Si(0)$,
i.e.\ we discard all diagrams containing a momentum-independent
self-energy part. 
\begin{table}
\begin{tabular}{|c||c|c|c|c|c|c|c|c|}
\hline
$L$ & 3 & 4 & 5 & 6 & 7 & 8 & 9 & 10 \\\hline
$n_L$ & 1 & 1 & 5 & 12 & 56 & 230 & 1262 & 7295 \\
\hline
\end{tabular}
\caption{\label{nd}Numbers of diagrams in low loop orders.}
\end{table}
The first non-zero perturbative coefficient arises at the three-loop
level, since the lowest momentum-dependent self-energy contribution
has two loops.
The coefficient is given by
\bea
\label{a3}
\lefteqn{a_3u_r=
\f{1}{Nu}{\cal R}
\rule[-14pt]{0pt}{34pt}
\bpi(30,0)(2,0)
\put(17,3){\circle{24}}
\put(17,3){\oval(24,8)}
\put(5,3){\circle*{2}}
\put(29,3){\circle*{2}}
\put(17,-9){\circle*{2}}
\epi}
\nn\\
&=&
\f{(N+2)u}{18}
\int\f{d^3k}{(2\pi)^3}\int\f{d^3p}{(2\pi)^3}
\int\f{d^3q}{(2\pi)^3}\f{1}{(k^2+r)^2}
\nn\\
&&\times
\left[\f{1}{(k+p)^2+r}-\f{1}{p^2+r}\right]\f{1}{[(p+q)^2+r](q^2+r)}
\nn\\
&=&
-\f{(1+\f{2}{N})\ln\f{4}{3}}{576\pi^2}u_r.
\eea
As already announced in Sec.\ \ref{fieldtheory}, the recursive
subtraction of zero-momentum self-energies in the integrands
eliminates all ultraviolet (UV) divergences and we
can work in three dimensions without regulator.

The numerical evaluation of Feynman diagrams is performed in
momentum space, using analytic results for one-loop subdiagrams
(the cases where one needs only one independent external momentum are
essentially trivial; see \cite{Ni} for the more complicated cases of
one-loop triangle and box diagrams) and the subtracted sunset diagram
(see Eq.\ (70) of \cite{BaBlHoLaVa2} and Eq.\ (A6) of \cite{BrRa1})
and carrying out the remaining integrations on the computer.
The necessary diagrams are among a larger set of diagrams that can
serve to obtain critical exponents in O($N$) field theories.
The results for diagrams through six loops in the renormalization
scheme employed here can be found in \cite{MutNi}.
The seven-loop diagrams have been computed by B.~Nickel in 1991
and used for the determination of critical exponents for $N=0,1,2,3$
in \cite{MurNi}.
Nickel's results for the diagrams through seven loops \cite{Nipc} are
displayed in Table \ref{results} together with their weights.
\begin{table*}
\begin{tabular}{|c|r|c|c|l||c|r|c|c|l|}
\hline
$L$-$n$ & \# from \cite{NiMeBa} & $w_{L-n}$ & $g_{L-n}$
& $(L/2{-}1)(8\pi)^LI_{L-n}$ &
$L$-$n$ & \# from \cite{NiMeBa} & $w_{L-n}$ & $g_{L-n}$
& $(L/2{-}1)(8\pi)^LI_{L-n}$
\\\hline
3-1 & 4-M3 & 1/6 & $g_{1}$ & $-0.575364144904$ &
7-20 & 125-M7 & 1/12 & $g_{16}$ & $\phantom{+}0.05799468900$\\\cline{1-5}
4-1 & 6-M4 & 1/4 & $g_{2}$ & $-0.817603121794$ &
7-21 & 120-M7 & 1/24 & $g_{16}$ & $\phantom{+}0.05689249345$\\\cline{1-5}
5-1 & 12-M5 & 1/8 & $g_{3}$ & $-0.950830436253$ &
7-22 & 117-M7 & 1/12 & $g_{16}$ & $\phantom{+}0.04248235153$\\
5-2 & 11-M5 & 1/12 & $g_{4}$ & $\phantom{+}0.063289032026$ &
7-23 & 93-M7 & 1/8 & $g_{16}$ & $\phantom{+}0.09439061787$\\
5-3 & 14-M5 & 1/36 & $g_{4}$ & $\phantom{+}0.034976834929$ &
7-24 & 102-M7 & 1/12 & $g_{16}$ & $\phantom{+}0.029781711859$\\
5-4 & 13-M5 & 1/4 & $g_{5}$ & $-0.522389299127$ &
7-25 & 121-M7 & 1/6 & $g_{16}$ & $\phantom{+}0.050866270685$\\
5-5 & 15-M5 & 1/4 & $g_{5}$ & $-0.810810317465$ &
7-26 & 141-M7 & 1/24 & $g_{16}$ & $\phantom{+}0.038380451088$\\\cline{1-5}
6-1 & 29-M6 & 1/16 & $g_{6}$ & $-1.032433915322$ &
7-27 & 118-M7 & 1/12 & $g_{16}$ & $\phantom{+}0.026414905961$\\
6-2 & 27-M6 & 1/8 & $g_{7}$ & $\phantom{+}0.092703296625$ &
7-28 & 91-M7 & 1/8 & $g_{16}$ & $\phantom{+}0.057057476340$\\
6-3 & 33-M6 & 1/12 & $g_{7}$ & $\phantom{+}0.045271743432$ &
7-29 & 99-M7 & 1/8 & $g_{17}$ & $-0.415203240084$\\
6-4 & 34-M6 & 1/24 & $g_{7}$ & $\phantom{+}0.070805748949$ &
7-30 & 139-M7 & 1/16 & $g_{17}$ & $-0.800600314818$\\
6-5 & 28-M6 & 1/6 & $g_{7}$ & $\phantom{+}0.062064397151$ &
7-31 & 104-M7 & 1/8 & $g_{18}$ & $-0.196062158014$\\
6-6 & 30-M6 & 1/4 & $g_{8}$ & $-0.558960605344$ &
7-32 & 127-M7 & 1/4 & $g_{18}$ & $-0.458493427150$\\
6-7 & 36-M6 & 1/4 & $g_{8}$ & $-0.831785654370$ &
7-33 & 110-M7 & 1/8 & $g_{19}$ & $-0.234481729265$\\
6-8 & 31-M6 & 1/8 & $g_{8}$ & $-0.406156736719$ &
7-34 & 130-M7 & 1/8 & $g_{19}$ & $-0.494415425222$\\
6-9 & 32-M6 & 1/2 & $g_{9}$ & $-0.316465247271$ &
7-35 & 108-M7 & 1/4 & $g_{18}$ & $-0.25405258372$\\
6-10 & 35-M6 & 1/2 & $g_{9}$ & $-0.535115380809$ &
7-36 & 109-M7 & 1/8 & $g_{18}$ & $-0.35337393251$\\
6-11 & 37-M6 & 1/4 & $g_{9}$ & $-0.667694545359$ &
7-37 & 126-M7 & 1/4 & $g_{18}$ & $-0.53390645444$\\
6-12 & 38-M6 & 1/6 & $g_{10}$ & $-0.21440337147$ &
7-38 & 135-M7 & 1/8 & $g_{18}$ & $-0.65804614052$\\\cline{1-5}
7-1 & 96-M7 & 1/32 & $g_{11}$ & $-1.085167325405$ &
7-39 & 113-M7 & 1/8 & $g_{20}$ & $-0.22686994808$\\
7-2 & 114-M7 & 1/216 & $g_{12}$ & $-0.002157445026$ &
7-40 & 138-M7 & 1/8 & $g_{20}$ & $-0.491439771$$^*$\\
7-3 & 92-M7 & 1/72 & $g_{12}$ & $-0.004646692505$ &
7-41 & 106-M7 & 1/4 & $g_{21}$ & $-0.238176055425$\\
7-4 & 98-M7 & 1/12 & $g_{13}$ & $\phantom{+}0.060934949636$ &
7-42 & 100-M7 & 1/2 & $g_{21}$ & $-0.323424309472$\\
7-5 & 95-M7 & 1/24 & $g_{13}$ & $\phantom{+}0.060934949636$ &
7-43 & 122-M7 & 1/4 & $g_{21}$ & $-0.418158972550$\\
7-6 & 137-M7 & 1/48 & $g_{13}$ & $\phantom{+}0.069176047800$ &
7-44 & 132-M7 & 1/4 & $g_{21}$ & $-0.542017150796$\\
7-7 & 116-M7 & 1/24 & $g_{13}$ & $\phantom{+}0.049619035559$ &
7-45 & 129-M7 & 1/4 & $g_{21}$ & $-0.625069918658$\\
7-8 & 90-M7 & 1/16 & $g_{13}$ & $\phantom{+}0.110191210376$ &
7-46 & 107-M7 & 1/1 & $g_{20}$ & $-0.17920527706$\\
7-9 & 112-M7 & 1/72 & $g_{12}$ & $-0.004476954704$ &
7-47 & 123-M7 & 1/2 & $g_{20}$ & $-0.31676596355$\\
7-10 & 115-M7 & 1/36 & $g_{12}$ & $-0.002517936782$ &
7-48 & 143-M7 & 1/4 & $g_{20}$ & $-0.35422809760$\\
7-11 & 89-M7 & 1/24 & $g_{12}$ & $-0.004609529361$ &
7-49 & 128-M7 & 1/2 & $g_{20}$ & $-0.32400906607$\\
7-12 & 94-M7 & 1/4 & $g_{14}$ & $\phantom{+}0.086255985441$ &
7-50 & 136-M7 & 1/4 & $g_{20}$ & $-0.49327378592$\\
7-13 & 119-M7 & 1/16 & $g_{14}$ & $\phantom{+}0.102504909593$ &
7-51 & 131-M7 & 1/2 & $g_{20}$ & $-0.41989663865$\\
7-14 & 134-M7 & 1/16 & $g_{14}$ & $\phantom{+}0.057209654200$ &
7-52 & 111-M7 & 1/4 & $g_{22}$ & $-0.1112555703$$^*$\\
7-15 & 103-M7 & 1/16 & $g_{15}$ & $-0.340755240980$ &
7-53 & 142-M7 & 1/2 & $g_{22}$ & $-0.16148138890$\\
7-16 & 101-M7 & 1/8 & $g_{15}$ & $-0.581418054856$ &
7-54 & 140-M7 & 1/4 & $g_{22}$ & $-0.15678153586$\\
7-17 & 97-M7 & 1/16 & $g_{15}$ & $-0.581418054856$ &
7-55 & 133-M7 & 1/2 & $g_{22}$ & $-0.2010875739$$^*$\\
7-18 & 124-M7 & 1/8 & $g_{15}$ & $-0.847760133200$ &
7-56 & 144-M7 & 1/2 & $g_{23}$ & $-0.11081591890$\\
7-19 & 105-M7 & 1/6 & $g_{16}$ & $\phantom{+}0.037007306205$&&&&&\\
\hline
\end{tabular}
\caption{\label{results}
Weights $w_{L-n}$, group factors $g_{L-n}$ and numerical
results for diagrams through seven loops ($^*=$ probable
error in the two final digits).
The explicit group factors can be found in Table \ref{groupfactors}.
The numerical data were provided by B.~Nickel and have
only partially been checked by the author.
The results through six loops can be found in \cite{MutNi}.
We also provide the diagram numbering according to \cite{NiMeBa}.
The contribution of each diagram to $\langle\phi^2\rangle$ is 
$\Rcal D_{L\mbox{-}n}
=(-u)^{L-1}r^{1-L/2}w_{L\mbox{-}n}g_{L\mbox{-}n}I_{L\mbox{-}n}$.}
\end{table*}
The different group factors are collected in Table \ref{groupfactors}.
\begin{table}
\begin{tabular}{|r|c|}
\hline
$n$&$g_n$\\\hline
1&$N(2+N)/3$\\
2&$N(2+N)(8+N)/27$\\
3&$N(2+N)\left(20+6N+N^2\right)/81$\\
4&$N(2+N)^2/9$\\
5&$N(2+N)(22+5N)/81$\\
6&$N(2+N)\left(48+24N+8N^2+N^3\right)/243$\\
7&$N(2+N)^2(8+N)/81$\\
8&$N(2+N)\left(56+22N+3N^2\right)/243$\\
9&$N(2+N)\left(60+20N+N^2\right)/243$\\
10&$N(2+N)(22+5N)/81$\\
11&$N(2+N)\left(112+80N+40N^2+10N^3+N^4\right)/729$\\
12&$N(2+N)^3/27$\\
13&$N(2+N)^2\left(20+6N+N^2\right)/243$\\
14&$N(2+N)^2(8+N)^2/729$\\
15&$N(2+N)\left(136+80N+24N^2+3N^3\right)/729$\\
16&$N(2+N)^2(22+5N)/243$\\
17&$N(2+N)\left(144+80N+18N^2+N^3\right)/729$\\
18&$N(2+N)\left(156+76N+11N^2\right)/729$\\
19&$N(2+N)\left(160+72N+10N^2+N^3\right)/729$\\
20&$N(2+N)\left(164+72N+7N^2\right)/729$\\
21&$N(2+N)\left(152+76N+14N^2+N^3\right)/729$\\
22&$N(2+N)(8+N)(22+5N)/729$\\
23&$N(2+N)\left(186+55N+2N^2\right)/729$\\
\hline
\end{tabular}
\caption{\label{groupfactors}
Group factors $g_n$ for Table \ref{results}.}
\end{table}
The resulting perturbative coefficients $a_L$ for $N=1,2,4$ are displayed
in Table \ref{acoeffs}.
\begin{table*}
\begin{tabular}{|c|l|l|l|}
\hline
$L$& \multicolumn{1}{c|}{$a_L$ for $N=1$}
& \multicolumn{1}{c|}{$a_L$ for $N=2$}
& \multicolumn{1}{c|}{$a_L$ for $N=4$} \\\hline
3
& $-1.51814\times10^{-4}$
& $-1.01209\times10^{-4}$
& $-7.59070\times10^{-5}$\\
4
& $\phantom{+}8.08989\times10^{-5}$
& $\phantom{+}2.99626\times10^{-5}$
& $\phantom{+}1.34832\times10^{-5}$\\
5
& $-5.88280\times10^{-5}$
& $-1.19872\times10^{-5}$
& $-3.17492\times10^{-6}$\\
6
& $\phantom{+}5.25790\times10^{-5}$
& $\phantom{+}5.85519\times10^{-6}$
& $\phantom{+}9.01670\times10^{-7}$\\
7
& $-5.45889\times10^{-5}$
& $-3.30467\times10^{-6}$
& $-2.93269\times10^{-7}$\\
\hline
\end{tabular}
\caption{\label{acoeffs}
Perturbative coefficients $a_L$ through seven loops for $N=1,2,4$.
$a_1=a_2=0$ for any $N$.}
\end{table*}

\section{Resummation and Results}
\label{resummation}
\begin{figure}[ht]
\begin{center}
\includegraphics[width=8cm,angle=0]{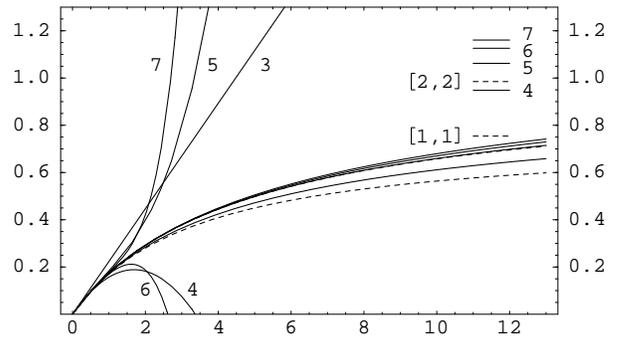}
\end{center}
\vspace{-15pt}
\caption{\label{fig1}
Results for $c_1(u_r)$ in VPT with fixed $\om'=0.805$
for $N=2$ for four through seven loops
and with $[1,1]$ and $[2,2]$ Pad\'{e} approximants (dashed lines; in the
range of the plot, the $[2,2]$ approximant is almost indistinguishable
from the five-loop VPT curve; the $[1,1]$ and $[2,2]$ approximants
use results through four and six loops, respectively).
For $u_r\rightarrow\infty$ one obtains the fixed-$\om'$
results for $c_1$, indicated by the line segments at the upper right.
Also indicated are the $u_r\ra\infty$ limits of the Pad\'{e}
approximants. 
The loop order rises from bottom to top for both the curves at the
right end of the plot and their limiting values.
Also shown are the unresummed perturbative results from three through
seven loops.}
\end{figure}

Being interested in the $u_r\ra\infty$ limit of $c_1(u_r)$, we need to
resum our perturbative series (\ref{pertsum}).
One of the simplest such resummations is the reexpansion in Pad\'{e}
approximants.
Since we are computing a finite quantity, we only consider diagonal
approximants $[n,n]$.
The $[1,1]$ and $[2,2]$ approximants use perturbative coefficients
through four and six loops, respectively.
They are plotted in Fig.\ \ref{fig1} together with their limiting
values $c_1=0.754$ for $[1,1]$ and $c_1=0.985$ for $[2,2]$.

We can do better than that, though.
First of all, diagonal Pad\'{e} approximants allow only to work
consistently through even loop orders.
Second, their large-$u_r$ behavior is
\beq
\label{larggeur}
c_1(u_r)=\al\sum_{m=0}^\infty f_mu_r^{-m\om'},
\eeq
with $\om'=1$.
However, the interactions cause the phase transition to be second order
with critical exponents of the O(2) universality class.
The leading class of corrections of a physical quantity that remains
finite in the critical limit are integer powers of $t^{\om\nu}$
\cite{phi4book,We,ZiPeVi}, where $t\equiv(T-T_c)/T_c$,
$\nu$ is the critical exponent of the correlation length and
$\om=\be'(g^*)$ in a renormalization group approach.
Since, in our renormalization scheme, the propagator obeys
$G(p=0)=1/r\propto t^{-\gamma}$, the leading corrections are integer
powers of $u_r^{-\om'}$ with
\beq
\label{omp}
\om'=\f{2\om}{2-\eta}.
\eeq
Here we have employed the universal scaling relation $\gamma=\nu(2-\eta)$,
where $\eta$ is the anomalous dimension of the critical propagator,
i.e.\ $G(r=0)\propto1/p^{2-\eta}$ in the small-$p$ limit.
Thus an expansion which correctly describes the leading corrections
to scaling \cite{We} has the form (\ref{larggeur}) with $\om'$ from
(\ref{omp}).

The ansatz (\ref{larggeur}) does not account for so-called confluent
singularities which cause the true large-$u_r$ expansion to also contain
other negative powers of $u_r$, which are subleading at least compared
to $u_r^{-\om'}$.
We can expect methods that can accommodate the leading
behavior in (\ref{larggeur}) correctly to converge faster to the
true result than methods having the wrong leading behavior as e.g.\
Pad\'{e} approximants or the LDE, the latter being used extensively for
the current problem \cite{SCPiRa1,SCPiRa2,KnPiRa,BrRa1,BrRa2}.
On the other hand, convergence will be slowed by the fact that we
do not make an ansatz reflecting the full power structure in $u_r$,
but the expansion (\ref{larggeur}) will try to mimic the neglected
subleading powers.

The alternating signs of the $a_L$ displayed in Table \ref{acoeffs}
suggest that the perturbation series for $c_1(u_r)$ is Borel summable.
In the context of critical phenomena, such series have been successfully
resummed using Kleinert's VPT.
Accurate critical exponents \cite{Kl3,Kl4,phi4book} and amplitude ratios
\cite{KlvdB} have been obtained.
For a truncated partial sum $\sum_{l=1}^La_lu_r^{l-2}$ of (\ref{pertsum}),
the method requires replacing
\bea
u_r^{l-2}
\!&\ra&\!
(t\uh)^{l-2}
\left\{1+t\left[\left(\f{\uh}{u_r}\right)^{\om'}-1\right]
\right\}^{-(l-2)/\om'}
\eea
(note that this is an identity for $t=1$),
reexpanding the resulting expression in $t$ through $t^{L-2}$, setting
$t=1$ and then optimizing in $\uh$, where optimizing is done in accordance
with the principle of minimal sensitivity \cite{Ste} and in practice means
finding appropriate stationary or turning points.
That is, we replace
\beq
\label{uuhat}
u_r^{l-2}
\ra
\uh^{l-2}\sum_{k=0}^{L-l}
\left(\ba{c}-(l-2)/\om'\\k\ea\right)
\left[\left(\f{\uh}{u_r}\right)^{\om'}-1\right]^k
\eeq
and optimize the resulting expression in $\uh$.
For $u_r\ra\infty$, we obtain the $L$-loop approximation of $f_0$,
\beq
\label{f0}
f_0^{(L)}={\rm opt}_{\uh}\left[\sum_{l=1}^La_l\uh^{l-2}\sum_{k=0}^{L-l}
\left(\ba{c}-(l-2)/\om'\\k\ea\right)(-1)^k\right].
\eeq
Results are only available starting at four loops, since two non-zero
perturbative coefficients are necessary for VPT to work.
E.g., in four-loop order the optimization yields
\beq
f_0^{(4)}={\rm opt}_{\uh}
\left[a_3\left(1+\f{1}{\om'}\right)\uh+a_4\uh^2\right]
=-\f{a_3^2\left(1+\f{1}{\om'}\right)^2}{4a_4}
\eeq
as the best attempt at determining $f_0$ and thus $c_1=\al f_0$.

\begin{figure}[ht]
\bigskip
\begin{center}
\includegraphics[width=8cm,angle=0]{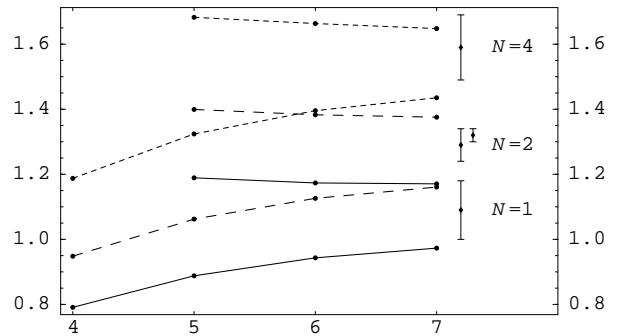}
\end{center}
\vspace{-15pt}
\caption{\label{fig2}
$c_1$ from VPT as a function of the number of loops $L$ for $N=1$ (solid),
$N=2$ (long dashes) and $N=4$ (short dashes).
In each case, the rising lines are using a fixed $\om'$, while for the
falling lines, $\om'$ is determined self-consistently.
For comparison, the three-dimensional MC results from \cite{KaPrSv} and
\cite{ArMo} for $N=2$ and from \cite{Su} for $N=1,4$ have been included.}
\end{figure}

For $N=2$, we have $\om=0.79\pm0.01$, $\eta=0.037\pm0.003$
(see, e.g., \cite{phi4book}) and therefore, using (\ref{omp}),
$\om'=0.805\pm0.011$.
We have determined $c_1(u_r)$ at loop orders $L=4,5,6,7$ for arbitrary
$u_r$ and plotted the results together with their large-$u_r$ limit
in Fig.\ \ref{fig1}.
Also plotted are the unresummed perturbative results.
It is instructive to see how VPT converts them to the smooth
function $c_1(u_r)$ and how, after resummation, the different loop
orders have apparent converging behavior for any $u_r$.
Table \ref{c1om} and Fig.\ \ref{fig2} display the limiting values $c_1$ for
fixed $\om'$ for $N=1,2,4$.
\begin{table*}
\begin{tabular}{|c||c|c|c||c|c|c||c|c|c||c|c|c|}
\hline
& \multicolumn{3}{c||}{$N=1$} & \multicolumn{3}{c||}{$N=2$} &
\multicolumn{3}{c||}{$N=4$} \\\hline
& fixed $\om'$ & \multicolumn{2}{c||}{s.-c.\ $\om'$}
& fixed $\om'$ & \multicolumn{2}{c||}{s.-c.\ $\om'$}
& fixed $\om'$ & \multicolumn{2}{c||}{s.-c.\ $\om'$}
\\\hline
$L$ & $\om'=0.804$ & $c_1$ & $\om'$ & $\om'=0.805$ & $c_1$ & 
$\om'$ & $\om'=0.803$ & $c_1$ & $\om'$ \\\hline
$4$ & $c_1=0.791$ & && $c_1=0.948$ &&& $c_1=1.188$ &&\\
$5$ & $c_1=0.888$ & $1.189$ & $0.6135$ & $c_1=1.062$ & $1.399$ & $0.6212$ 
& $c_1=1.324$ & $1.683$ & $0.6369$ \\
$6$ & $c_1=0.943$ & $1.173$ & $0.6314$ & $c_1=1.126$ & $1.383$ & $0.6381$ 
& $c_1=1.396$ & $1.663$ & $0.6528$ \\
$7$ & $c_1=0.973$ & $1.171$ & $0.6419$ & $c_1=1.161$ & $1.376$ & $0.6504$ 
& $c_1=1.435$ & $1.648$ & $0.6681$ \\
\hline
\end{tabular}
\caption{\label{c1om}$c_1$ with fixed $\om'$ and $c_1$ and $\om'$
for self-consistent $\om'$ for $N=1,2,4$ through loop order $L=7$.}
\end{table*}
From \cite{phi4book} we take $\om=0.79\pm0.01$, $\eta=0.035\pm0.003$
and thus $\om'=0.804\pm0.011$ for $N=1$ and $\om=0.79\pm0.01$,
$\eta=0.0335\pm0.0030$ and thus $\om'=0.803\pm0.011$ for $N=4$.
The uncertainties of the $\om'$ contribute negligibly to the
errors of the final results that we obtain in (\ref{c1n1})-(\ref{c1n4})
below for $c_1$.

There is also the possibility to determine $\om'$ self-consistently
\cite{Kl3,pibook,phi4book}.
It relies on the fact that assuming a behavior of $c_1(u_r)$ as in
(\ref{larggeur}), the quantity $d\ln c_1(u_r)/d\ln u_r$ has an expansion
of the same type with the same $\om'$ as $c_1(u_r)$, but with a vanishing
large-$u_r$ limit (i.e.\ its large-$u_r$ expansion starts out with
$f_0=0$).
$\om'$ is tuned such that the value VPT gives for
$d\ln c_1(u_r)/d\ln u_r$ is zero in a given loop order.
This $\om'$ is then used as an input for the determination of the
approximation of $c_1(u_r)$ at the same loop order.
We have used this method to determine $c_1$ at loop orders
$L=5,6,7$ for $N=1,2,4$ with the results for $c_1$ and $\om'$ given
in Table \ref{c1om} and displayed in Fig.\ \ref{fig2}.
The $\om'$ appear to approach their correct values at best very
slowly.
This may be interpreted as an indication of the presence of confluent
singularities as described above.
The self-consistently determined $\om'$ values then try to mimic the
true mixture of powers.
One may speculate that this causes the method of self-consistently
determined $\om'$ to be ultimately superior than using the correct
leading power $\om'$, whenever the power structure is more complicated
than that in (\ref{larggeur}).

It appears that, in our treatment, for each $N$ the correct value for
$c_1$ is approached from below by the $c_1$ values for fixed $\om'$
and from above for self-consistently determined $\om'$.
We therefore take the difference of the results as a conservative
estimate for the full error bar and obtain at seven-loop
order
\bea
\label{c1n1}
N=1:&&c_1=1.07\pm0.10,
\\
\label{c1n2}
N=2:&&c_1=1.27\pm0.11,
\\
\label{c1n4}
N=4:&&c_1=1.54\pm0.11.
\eea

\section{Discussion}
\label{discussion}
\begin{figure}[ht]
\bigskip
\begin{center}
\includegraphics[width=8cm,angle=0]{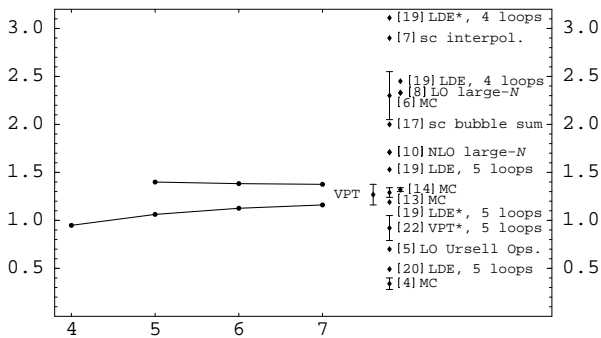}
\end{center}
\vspace{-15pt}
\caption{\label{fig3}
Comparison of $c_1$ from VPT as a function of the number of loops $L$ for
$N=2$ with most results from other sources.
Upper dots: self-consistent (sc) $\om'$.
Lower dots: $\om'=0.805$.
The label LDE$^\star$ indicates use of a resummation method for
accelerated convergence. 
The label VPT$^\star$ indicates the inclusion of a non-zero one-loop
term in Ref.\ \cite{Kl1}, which is absent in the present treatment,
labeled by VPT.
Not shown are other approximate results of \cite{BaBlHoLaVa2},
the renormalization group results $c_1=4.7$ \cite{Sto} and $c_1=3.42$
\cite{Al} and the canonical ensemble result $c_1=-0.93$ \cite{WiIlKr}.}
\end{figure}

There is a plethora of results for the parameter $c_1$ describing the
leading deviation of the BEC temperature due to a small repulsive
interaction.
A comparison of our results through seven loops with most other results
found in the literature is given in Fig.\ \ref{fig3}.
Let us have a closer look concerning their reliability.

The large-$N$ results \cite{BaBlZi,ArTo} move towards the region of
(\ref{c1n2}), but their convergence behavior is unclear.
The results obtained from a functional Keldysh formalism \cite{Sto},
the leading-order Ursell Operator method \cite{HoGrLa},
the self-consistent method of \cite{BaBlHoLaVa1},
the canonical ensemble \cite{WiIlKr}
and the renormalization group \cite{Al}
come without any next order or error bar.
The low MC value \cite{GrCeLa} has found explanations in
\cite{ArMo,HoBaBlLa,BaBlHoLaVa2}.
The setup of the MC simulation of \cite{HoKr} has been criticized in
\cite{ArMo}.
The LDE and LDE$^\star$ results of
\cite{SCPiRa1,SCPiRa2,KnPiRa,BrRa1,BrRa2} show no apparent converging
behavior.
Therefore, the closeness of the five-loop results of \cite{KnPiRa} to the
three-dimensional MC data and our VPT result appears somewhat accidental.
Use of the LDE in the context of field theoretic calculations with
anomalous approach-to-scaling exponents has been criticized on
general grounds in \cite{HaKl}.

The VPT treatment of \cite{Kl1} has inspired \cite{Ka} and its extension
to seven loops discussed in the current work.
However, as argued at the end of Sec.\ \ref{fieldtheory}, the choice
of a massless free reference propagator in \cite{Kl1} even before tuning
the full propagator to the critical limit causes a non-zero one-loop
result that strongly influences resummation.

Our result (\ref{c1n2}) is in agreememt with the apparently most
reliable other sources available, the MC results from the
three-dimensional theory,
$c_1=1.29\pm0.05$ \cite{KaPrSv} determined by Kashurnikov, Prokof'ev and
Svistunov and $c_1=1.32\pm0.02$ \cite{ArMo} determined by Arnold
and Moore.

MC results for $N=1$ and $N=4$ are also available.
X.~Sun has computed $\De\langle\phi^2\rangle_{\rm crit.}/u$ for these
cases \cite{Su} which translate to $c_1=1.09\pm0.09$ for $N=1$ and
$c_1=1.59\pm0.10$ for $N=4$.
Our values (\ref{c1n1}) and (\ref{c1n4}) agree well with these results,
which is also illustrated in Fig.\ \ref{fig2}.
It appears exceedingly unlikely that the agreement between the completely
different approaches of MC and VPT for all three values of $N$ is
accidental.

\section*{Acknowledgments}
The author is extremely grateful to B.~Nickel for providing the
results of his seven-loop calculations, without which the extension
of our work through seven loops would not have been possible. 
He thanks H.~Kleinert for important suggestions, many
discussions,  and a careful reading of the manuscript.
He is grateful to P.~Arnold for many helpful communications and for
suggesting the extension of our work to $N=1,4$ to allow for comparison
with the results of \cite{Su}.
He thanks E.~Braaten for many valuable suggestions.
Useful communications with J.~Andersen, M.~Holzmann, F.~Lalo\"e, D.~Murray,
M.~Pinto and R.~Ramos are also acknowledged.


\begin{thebibliography}{99}
\bibitem{Kl1res}
The final result $c_1=1.14\pm0.11$ of \cite{Kl1} was obtained by also
applying VPT with self-consistent $\om'$ to scheme III of \cite{BrRa1}
(as first done in \cite{Ka}) through five loops (albeit with the imprecise
numerical values for the integrals given in \cite{BrRa1})  and combining
the result with $c_1=0.92\pm0.13$.
\end{thebibliography}
\end{document}